\documentclass[aps,twocolumn]{revtex4}
\usepackage{newlfont}
\usepackage{amssymb}
\usepackage{amsfonts}
\usepackage{amsmath}

\usepackage{graphicx}
\usepackage{bm}

\usepackage{graphicx}
\usepackage{epsfig}
\usepackage{newlfont}
\usepackage{amssymb}
\usepackage{amsfonts}
\usepackage{amsmath}
\usepackage{graphicx}
\usepackage{bm}

\usepackage{amsthm}

\begin{document}



\title{Channel Capacities versus Entanglement Measures in Multiparty Quantum States}

\author{Aditi Sen(De) and Ujjwal Sen}

\affiliation{Harish-Chandra Research Institute, Chhatnag Road, Jhunsi, Allahabad 211 019, India}

\begin{abstract}
For quantum states of \emph{two} subsystems,  entanglement measures are related to capacities of 
communication tasks -- highly entangled states give higher capacity of transmitting classical as well as quantum information. However, 
we show that 
this is no more the case in general: quantum capacities of multi-access channels, motivated by 
communication in quantum networks, 
do not have any relation with 
genuine multiparty entanglement measures. 
Along with revealing the structural richness of multi-access channel capacities,  
this gives us a tool to classify multiparty quantum states from the perspective of its usefulness in quantum 
networks, which cannot be visualized by known multiparty entanglement measures. 
\end{abstract}

\maketitle

\section{Introduction and Main Results}
Understanding quantum entanglement \cite{HHHH-RMP} has 
been one of the key features in the development of the science of quantum information 
 \cite{ref-boi}. 
Applications of quantum information had started off in the fields of 
communication, cryptography, computation, and thermodynamics \cite{ref-boi}, and 
has since diffused into diverse areas such as condensed matter physics, ultra-cold gases, and statistical mechanics  
 \cite{ref-reviews}. 
Measuring and detecting entanglement of the quantum states appearing in different physical 
situations has been the cornerstones of the development in these directions. 
It has therefore been very important to propose 
entanglement measures of general quantum states of systems consisting of more than one
subsystem, and there is a thriving 
industry of such proposals (see \cite{ref-measures,HHHH-RMP} and references therein). 
However, the main progress in the theory 
of entanglement measures, and its detection, has been in the case when the physical 
system consists of only two subsystems. 
This has been a major handicap in using entanglement as an instrument for handling many-body physics 
systems like ultra-cold atomic states, where the 
majority, if not all, of the quantum states involved are of multiparty systems, i.e. a physical 
system consisting of more than two subsystems. Understanding 
multiparty quantum entanglement is therefore a distinct necessity to a large 
portion of physics of our times. 

One of the main reasons for the current interest in quantum information is its potential for 
revolutionizing future communication systems. It is therefore hard to overestimate the importance of 
capacities of quantum communication channels \cite{ref-capa}. 
Again the main progress in research in this area has been for quantum channels between a single sender and 
a single receiver, while multi-access channels clearly have more commercial viability. Moreover, a competent functioning of future 
quantum computers \cite{ref-boi} will require efficient communication of quantum information between its different parts.

The archetypical quantum channels are bipartite quantum states used as dense coding \cite{ref-dense} and teleportation \cite{ref-tele} 
channels. They are channels respectively for transmitting classical and quantum information, and 
form the basis of most
quantum channels. 
If a \emph{pure} bipartite quantum state \(\left|\Psi\right\rangle_{AB}\) (\(\in \mathbb{C}^d \otimes \mathbb{C}^d\)) 
is shared between Alice (\(A\)) and Bob (\(B\)), 
it can be used as a quantum channel to perform dense coding, by which classical information can be sent, for example, by Alice to Bob, with the capacity
(measured in \emph{bits}) being 
 \({\cal C}_{\mbox{\scriptsize{classical}}} \left(\left|\Psi\right\rangle\right)= \log_2 d + S\left(\varrho_L\right)\) \cite{ref-dense, amadersobar},
where \(\varrho_L\) is the local density matrix of state \(\left|\Psi\right\rangle_{AB}\), and \(S(\cdot)\) is the von Neumann
entropy of the argument. Similarly, the same quantum state 
\(\left|\Psi\right\rangle_{AB}\) can be used as a quantum channel to convey quantum information from \(A\) to \(B\), with the 
capacity  (measured in \emph{qubits})
 being \({\cal C}_{\mbox{\scriptsize{quantum}}} \left(\left|\Psi\right\rangle\right)=  S\left(\varrho_L\right)\) \cite{ref-tele, ref-capa, ref-pure-conc}.
%
%
Entanglement of a bipartite pure quantum state \(\left|\Psi\right\rangle_{AB}\) is, for most purposes,
 the von Neumann entropy of a local subsystem, i.e.  
\(E\left(\left|\Psi\right\rangle\right) = S\left(\varrho_L\right)\) \cite{ref-pure-conc}.

Clearly, higher entanglement for a pure quantum state implies higher capacities for both the classical and quantum instances, in the case
of a single sender and a single receiver. 
%
%
Here we find that a generalization of this behavior is not mirrored in the multiparty case. More precisely, 
we find quantum capacities 
of four-party quantum states that are motivated by considering distillation protocols in 
multiparty quantum networks, and show that their values are not correlated with those 
of a measure of genuine four-party entanglement. The measure
of genuine four-party entanglement that  we use here is a generalization of the ``geometric measure
of entanglement'' (GM) \cite{ref-gyamiti}, and we call it the ``generalized geometric measure'' (GGM). 
As an important by-product, we obtain a \emph{computable} measure of genuine multiparty entanglement, 
which can potentially have the same usefulness in the multiparty case, as the logarithmic negativity \cite{ref-logneg} has in the bipartite situation.
We also provide bounds on the capacities defined that help us in their understanding as well as their evaluation in a variety of paradigmatic classes of 
multipartite quantum states.

\section{The multi-access capacities}
Let us begin by defining the multi-access capacities that we will deal with, and by considering their quantum computational significance.
We will define two such quantities, both of which are given from the perspective of quantum networks. 
Although the definitions, and the subsequent propositions, will be given only for four-party systems, their generalizations to more
parties (or for three parties) are straightforward.
The first quantity is \emph{maximal assisted remote singlet production}
 (\({\cal C}_a\)), and defined for a 
single copy of a four-party pure quantum state, \(\left|\psi\right\rangle\), shared between 
Alice (\(A\)), Bob (\(B\)), Claire (\(C\)), and Danny (\(D\)),
as the maximal probability with which a single copy of the singlet state, 
\(\left|\psi^-\right\rangle = \left( \left|01\right\rangle - \left|10\right\rangle  \right)/\sqrt{2}\)
(\(\left|0\right\rangle\) and \(\left|1\right\rangle\) are mutually orthonormal),
can be prepared at \(CD\), by using an additional 
resource of a singlet state shared between Alice and Bob, and by using local 
quantum operations and classical communication
(LOCC) between Alice, Bob, Claire, and Danny. \({\cal C}_a\) therefore measures
the amount of entanglement that can be transferred from Alice and Bob to Claire and Danny, when Alice and Bob are 
assisted by an additional singlet. It is therefore natural to multiply this quantity by the entanglement (\(E\)) value of 1 ebit of the singlet
state, and express the capacity in ebits. 
If the state is not symmetric with respect to interchange of the parties, we define \({\cal C}_a\) as the 
maximum 
of the transfer probabilities corresponding to all possible permutations of the parties.

The other quantity is \emph{maximal unassisted remote singlet production} (\({\cal C}_{ua}\)), 
and has exactly the same definition as \({\cal C}_a\), but without the additional singlet assistance. 
These quantities, or their generalized versions for large quantum networks, are important elements 
in quantum computational setups, e.g. in the Knill-Laflamme-Milburn model of quantum computation \cite{ref-KLM}, or 
in the cluster state model of quantum computation \cite{eita-cluster} (see also \cite{eita-KLM-Nielsen}).


The multi-access capacities \( {\cal C}_{ua}\) and  \( {\cal C}_a \) can be shown to be  monotonically decreasing under LOCC between the four observers. 
More importantly, we have the following results.

\textbf{Proposition C1.} \( {\cal C}_{ua} \leq {\cal C}_a \leq p_{\mbox{\scriptsize{maxmin}}}^s\),
where \(p_{\mbox{\scriptsize{maxmin}}}^s\) is defined as follows. Consider the set of four quantities 
\(\left\{p_{\max}^{i: \mbox{\scriptsize{rest}}} | i=A,B,C,D \right\}\),
where e.g. \(p_{\max}^{C:ABD}\) 
is the maximum probability of obtaining a singlet between 
Claire and the other observers (who are at the same location), and where \(A\), \(B\), \(C\), and \(D\) share the 
quantum state \(\left|\psi\right\rangle\).
Choose all six pairs from the set \(\left\{p_{\max}^{i: \mbox{\scriptsize{rest}}}\right\}\), find the minimum for each pair, and 
then the maximum  of these six minima is \(p_{\mbox{\scriptsize{maxmin}}}^s\). \\
\emph{Proof.} The definitions of  \( {\cal C}_{ua}\) and \( {\cal C}_a\) imply the first inequality. 
Now, \(p_{\max}^{C:ABD}\left( \left|\psi\right\rangle_{ABCD} \right) = 
p_{\max}^{C:ABD}\left( \left|\psi\right\rangle_{ABCD} \otimes \left|\psi^-\right\rangle_{AB} \right)\), as 
adding a local ancilla (local with respect to the \(C:ABD\) split) cannot change an LOCC monotone \(p_{\max}\). Further, 
\(p_{\max}^{C:ABD}\left( \left|\psi\right\rangle_{ABCD} \otimes \left|\psi^-\right\rangle_{AB} \right) 
\geq p_{\max}^{AB \rightarrow CD} \left( \left|\psi\right\rangle_{ABCD} \otimes \left|\psi^-\right\rangle_{AB} \right)\),
where \(p_{\max}^{AB \rightarrow CD}\) is the probability that \(A\) and \(B\) can create a singlet between
\(C\) and \(D\), when all four parties, at separated locations, share the quantum state in the argument. 
%
%
This is because the probability of a singlet being prepared between \(C\) and \(D\) by LOCC between all four observers, cannot exceed the
corresponding probability when \(A\), \(B\), and \(D\) are together. Similar relations hold when Claire is replaced by Danny,
and 
so we have 
\(p_{\max}^{AB \rightarrow CD} \left( \left|\psi\right\rangle_{ABCD} \otimes \left|\psi^-\right\rangle_{AB} \right)
\leq 
\min \left\{ p_{\max}^{j:ABk}\left( \left|\psi\right\rangle_{ABCD} \right) |j,k=C,D; j\ne k \right\}
\).
Taking the maximum, over all possible permutations of the four parties, in the preceding inequality, we obtain
the second inequality in the proposition. 
\hfill 
\(\blacksquare\)


\textbf{Proposition C2.} 
\({\cal C}_{ua} \leq p_{\mbox{\scriptsize{maxmin}}}^d\), where
\(p_{\mbox{\scriptsize{maxmin}}}^d\) is defined as follows. Consider the set of three quantities
\(\left\{ p_{\max}^{AC:BD}, p_{\max}^{AD:BC}, p_{\max}^{AB:CD}   \right\}\), where e.g. 
\(p_{\max}^{AC:BD}\) is the maximum probability of obtaining a singlet in the \(AC:BD\) bipartite split, and where \(A\), \(B\), \(C\), and \(D\) share the 
quantum state \(\left|\psi\right\rangle\). 
Choose all three pairs from the set, and find the minimum for each set. 
   \(p_{\mbox{\scriptsize{maxmin}}}^d\) is the maximum of these minima.\\
\emph{Proof.} Suppose that Alice and Claire are together, and so are Bob and Danny. The probability of 
preparing a singlet state in the \(AC:BD\) bipartite split must be greater than or equal to the 
corresponding quantity in the situation when all four parties are at 
separate locations, and the singlet is to be prepared between Claire and Danny. That is,
\(p_{\max}^{AC:BD}\left(\left|\psi\right\rangle_{ABCD}\right) \geq p_{\max}^{AB \rightarrow CD}\left(\left|\psi\right\rangle_{ABCD}\right)\).
A similar inequality holds when Alice and Bob change sides, i.e. 
\(p_{\max}^{AD:BC}\left(\left|\psi\right\rangle_{ABCD}\right) \geq p_{\max}^{AB \rightarrow CD}\left(\left|\psi\right\rangle_{ABCD}\right)\), so that we have
\(p_{\max}^{AB \rightarrow CD}\left(\left|\psi\right\rangle_{ABCD}\right) \leq 
\min\left\{p_{\max}^{AC:BD}\left(\left|\psi\right\rangle_{ABCD}\right), p_{\max}^{AD:BC}\left(\left|\psi\right\rangle_{ABCD}\right)\right\} \).
Taking a maximum, over all possible permutations of the four observers, of the preceding inequality proves the inequality in the proposition. 
\hfill \(\blacksquare\)


\section{The generalized geometric measure} These capacities, motivated by quantum networks,  will be compared with a measure of genuine four-party 
entanglement measure, GGM, which we now define. Consider a four-party pure quantum state
\(\left|\psi\right\rangle\), and  let 
\(\Lambda_{\max}\left( \left|\psi\right\rangle \right) = \max \left| \left\langle \phi | \psi \right\rangle \right|\),
where 
the maximum is over all four-party pure quantum 
states \(\left|\phi\right\rangle\) that are not genuinely four-party entangled.
An \(n\)-party pure quantum state is said to be genuinely \(n\)-party entangled, if 
it is not a product across any bipartite partition. The GGM of \(\left|\psi\right\rangle\)
is defined as 
\({\cal E} \left( \left|\psi\right\rangle \right) = 1- \Lambda^2_{\max}\left( \left|\psi\right\rangle \right)\). 
Note that \(\Lambda_{\max}\) quantifies the closeness of the state \(\left|\psi\right\rangle\) to all pure quantum states that are 
\emph{not} genuinely multiparty entangled. 
Generalization to arbitrary number of parties is straightforward. 
The definition is motivated by the GM, introduced in \cite{ref-gyamiti}, in which the maximization in \(\Lambda_{\max}\) is only 
over pure states that are product over every bipartite partition. [We denote the GM of a quantum state \(\left|\psi\right\rangle\) as 
\({\cal E}_G \left( \left|\psi\right\rangle \right)\).]
Clearly, \({\cal E}\) is vanishing for all pure states that are not genuine multiparty entangled, and non-vanishing for others.
We will now show that this measure is computable (for an arbitrary number of 
parties), and that it is indeed a monotonically decreasing quantity under LOCC.


\textbf{Proposition E1.} The generalized geometric measure can be written in closed (computable) form for all multipartite pure quantum states.\\
\emph{Proof.} We provide the proof for four-party states, the other cases being similar. 
The maximization in \(\Lambda_{\max}(|\psi\rangle_{ABCD}) = \max_{|\phi\rangle_{ABCD}} |\langle \phi | \psi \rangle|\) 
is over all pure quantum states \(|\phi\rangle_{ABCD}\) that are not genuinely multiparty entangled. 
The square of \(\Lambda_{\max}(|\psi\rangle_{ABCD})\) can therefore 
be interpreted as the Born probability of some outcome in a quantum measurement on the state \(|\psi\rangle\). 
However, since entangled measurements cannot be worse than the product ones for any set of subsystems, we have that the only measurements 
that we need to consider for the maximization are the ones in the 
single-party versus rest, and in the two-parties versus remaining-two splittings. Next we note that, if e.g. the maximization 
in \(\Lambda_{\max}(|\psi\rangle_{ABCD})\) is performed over all states that are product in the \(A:BCD\) split, the result is the 
maximal Schmidt coefficient, \(\lambda_{A:BCD}\), of the state \(|\psi\rangle_{ABCD}\), when written in the \(A:BCD\) split. Similar expressions hold for the other 
splittings. Therefore, we have that the GGM of \(|\psi\rangle\) is given by 
\({\cal E}(|\psi\rangle) = 1 - \max \{\lambda^2_{i:\mbox{\scriptsize{rest}}}, \lambda^2_{ij:\mbox{\scriptsize{rest}}}| i, j=A,B,C,D; i \ne j\}\).
%
%
\hfill \(\blacksquare\)

\textbf{Remark.} The closed form of the GGM may induce one to redefine the GGM in terms of R{\'e}nyi entropies of the Schmidt coefficients, especially for 
statistical mechanical applications. The current definition is akin to the case when the R{\'e}nyi parameter tends to infinity, sometimes referred to as 
the min-entropy.

\textbf{Proposition E2.} The generalized geometric measure is monotonically decreasing under LOCC. \\
\emph{Proof.} The proof follows from 
the fact that the \(\lambda\)'s involved in the 
closed form of the GGM, as derived in the proof of Proposition E1,
are all increasing under LOCC \cite{Vidal-Nielsen}. 
 \hfill  \(\blacksquare\)

\section{Applications and establishing the remaining results} 
With these results in hand, we now move to apply them to different classes of quantum states. 
As stated in the introduction, our main motivation is to study the defined quantum capacities from the perspective of quantum computational networks. 
In line with that, we begin by considering two classes of multiparty quantum states that have been found to be useful in several quantum informational and 
computational tasks.

\subsection{Case I: Generalized GHZ}
 A very important class of states, with several informational and computational applications,
is that of generalized Greenberger-Horne-Zeilinger states \cite{GHZ},
\(\left|\mbox{GHZ}_\alpha\right\rangle_{ABCD} = \alpha |0\rangle^{\otimes 4} + \beta |1\rangle^{\otimes 4}\)
(with \(|\alpha| \geq |\beta|\)), shared between the four observers. 
By Proposition C1,  \({\cal C}_{ua} \leq {\cal C}_a \leq 2 |\beta|^2\).
Supposing now that  measurements in the \(\left\{ \left|+\right\rangle, \left|-\right\rangle  \right\}\) basis, where
 \(|\pm\rangle = (|0\rangle \pm |1\rangle)/\sqrt{2}\), are carried out at both \( A\) and \(B\), and the 
resulting pure state at \(CD\), corresponding to each set of measurement results at \(A\) and \(B\), is  
LOCC-transformed to the singlet state, we obtain that \({\cal C}_{ua} \geq 2 |\beta|^2\) \cite{Vidal-Nielsen}. 
Therefore, \({\cal C}_{ua} = {\cal C}_a = 2 |\beta|^2\). For the generalized GHZ state \(\left|\mbox{GHZ}_\alpha\right\rangle_{ABCD}\), 
the GGM and GM coincide and are equal to \(|\beta|^2\). The GM is found by some algebra, while the GGM is found by using Proposition E1.
Therefore, for the GHZ state (generalized GHZ state for \(\alpha = \beta = 1/\sqrt{2}\)), the capacities are both 
unit ebits,
and the GGM and GM are both 
equal to one-half.

\subsection{Case II: Cluster states} From the point of view of quantum computational networks, the 
cluster states have acquired great significance \cite{eita-cluster}. The cluster state for four observers is 
\(|C\rangle_{ABCD}= \left( \left|0000\right\rangle + \left|0011\right\rangle + \left|1100\right\rangle- \left|1111\right\rangle \right)\).
It is a non-symmetric state. A trivial upper bound on the capacities is \({\cal C}_{ua} \leq {\cal C}_a \leq 1\). 
However, Alice and Bob can make measurements in the \(\left\{ \left|0\right\rangle, \left|1\right\rangle\right\}\) basis, and corresponding to every 
outcome, the state at the remaining parties turns out to be local unitarily equivalent to the singlet state. Therefore, we have that the unassisted 
capacity \({\cal C}_{ua}\) is 1 ebit, so that \({\cal C}_{ua} = {\cal C}_a = 1\). By explicit algebra, the GM for this state is \(3/4\), while the GGM is 
\(1/2\). [Let us mention here that the state \(\left|\chi\right\rangle_{ABCD} = \frac{1}{2\sqrt{2}}(|00\rangle (|00\rangle - |11\rangle) + |11\rangle (|00\rangle + |11\rangle)
- |01\rangle (|01\rangle - |10\rangle) + |10\rangle (|01\rangle + |10\rangle))\) of Ref. \cite{Yeo} has exactly the same values for 
\({\cal C}_{ua}\), \({\cal C}_{a}\), \({\cal E}\), and \({\cal E}_{G}\),
as the cluster state. However, the states are different, as can be seen by looking at the their entanglements in the \(AB:CD\) split.]

\subsection{Bipartite versus multipartite}  In the case of a single sender and a single receiver, we have seen that  the channel  capacities are 
consistently correlated with entanglement measures. Precisely, for two bipartite pure states \(\left|\Psi\right\rangle\) and 
\(\left|\Phi\right\rangle\), if \(E\left( \left|\Psi\right\rangle \right) = E\left( \left|\Phi\right\rangle \right) + \epsilon\) 
for some positive \(\epsilon\), then \({\cal C}\left( \left|\Psi\right\rangle \right) = {\cal C} \left( \left|\Phi\right\rangle \right) + 
\delta\) for some positive \(\delta\), where \({\cal C}\) is either \({\cal C}_{\mbox{\scriptsize{classical}}}\)
or \({\cal C}_{\mbox{\scriptsize{quantum}}}\).
In the case of multi-access quantum channels, we see 
that 
both the assisted and unassisted quantum capacities
(\({\cal C}_{a}\) and \({\cal C}_{ua}\) respectively) are unity 
for the GHZ state as well as for the cluster state, while
the geometric measure of entanglement returns different values for the two states. 
The GM however is not a measure of \emph{genuine} multiparty entanglement. [Indeed, it was defined by its inventors from quite a 
different perspective.]
The GGM, which \emph{is} a measure of genuine multiparty entanglement,
 of these two states are however equal, and so we regain the rather comfortable picture that is true in the case of a single sender and a single 
receiver. We will soon find that this simple picture to not hold in the case of 
multi-access channels. In any case, the fact that the picture does hold in certain cases also in the multi-access domain, especially 
in instances that are important from a quantum networks perspective, is still satisfying.
Let us now continue with our case studies.

\subsection{Case III: Generalized W} The next class of states that we handle is that of generalized W \cite{W-state} (see also \cite{amader-W}), defined as 
\(\left|\mbox{W}_{abcd}\right\rangle_{ABCD}= a|0001\rangle + b|0010\rangle + c|0100\rangle + d|1000\rangle\) (with 
\(|a| \geq |b| \geq |c| \geq |d|\)), and is another class of states with interesting conceptual and practical utilities in quantum information. This is 
also a non-symmetric state. By Proposition C1, \({\cal C}_{ua} \leq {\cal C}_a \leq p_{\mbox{\scriptsize{maxmin}}}^s= 2|b|^2\). 
Measuring in the \(\left\{ |0\rangle, |1\rangle \right\}\) basis at both \(A\) and \(B\), we find that \(|00\rangle_{AB}\)
clicks with probability \(p_{00}^{W_{abcd}}= |a|^2 + |b|^2\), and creates the state \(\left(a|01\rangle + b|10\rangle\right)/\sqrt{|a|^2 + |b|^2}\) at \(CD\).  
The latter can be locally transformed to the singlet state with probability \(2|b|^2/\sqrt{|a|^2 + |b|^2}\) \cite{Vidal-Nielsen}. The 
other clicks at \(A\) and \(B\) always
produces a product state. Therefore, \({\cal C}_{ua} \geq p_{00}^{W_{abcd}} \times 2|b|^2/\sqrt{|a|^2 + |b|^2} = 2|b|^2\), whereby
\({\cal C}_{ua} = {\cal C}_a = 2|b|^2\). By using Proposition E1, the GGM of the generalized W state is given by 
\({\cal E}\left(\left|\mbox{W}_{abcd}\right\rangle\right) = |d|^2\). 
So for the W state (generalized W state for 
\(a=b=c=d = 1/2\)), the capacities are both one-half ebits, and \({\cal E}\left(\left|\mbox{W}\right\rangle\right) = 1/4\).
[Also, \({\cal E}_G\left(\left|\mbox{W}\right\rangle\right) = 37/64 = 0.578125\).]

\subsection{Case IV: \(\mbox{W}_2\)} Let us now consider the state 
\(\left|\mbox{W}_2\right\rangle_{ABCD}= (|0011\rangle + |0110\rangle + |1100\rangle + |1001\rangle + |0101\rangle + |1010\rangle)/\sqrt{6}\).
Of course we have \({\cal C}_{ua} \leq {\cal C}_a \leq 1\). Now, in the assisted case, we can use the singlet assistance to teleport \cite{ref-tele, eita-swapping}
the \(B\)-part of the \(\mbox{W}_2\) state to \(A\) (or vice-versa). Subsequently, we make a measurement in the Bell basis \cite{eita-Bell-basis} at \(AB\), which 
creates a state at \(CD\), that is local unitarily equivalent to the singlet. Therefore, we have \({\cal C}_a \left( \left| \mbox{W}_2 \right\rangle \right) =1\). 
To find the unassisted capacity,
we will turn to Proposition C2, to find that \({\cal C}_{ua} \leq 2/3\). Suppose now that both Alice and Bob measure in the 
\(\left\{ |0\rangle, |1\rangle \right\}\) basis. The \(|00\rangle_{AB}\) and \(|11\rangle_{AB}\) outcomes at \(AB\) produce 
product states at \(CD\). However, both the \(|01\rangle_{AB}\) and the \(|10\rangle_{AB}\) outcomes at \(AB\) produce the state 
\(\left( \left|01\right\rangle + \left|10\right\rangle\right)/\sqrt{2}\) at \(CD\), each with probability \(1/3\). Consequently, we have
\({\cal C}_{ua} \geq 2/3\), so that \({\cal C}_{ua} \left( \left| \mbox{W}_2 \right\rangle \right) =2/3\). 
The GGM for the \(\mbox{W}_2\) state is \(1/3\). [\({\cal E}_{G} \left( \left| \mbox{W}_2 \right\rangle \right) =5/8 = 0.625\).]

\subsection{Case V: Two singlets} In this case, which we denote by \(\left|SS\right\rangle\), any two observers share a singlet, and the other two share another singlet.
This is a non-symmetric case. The unassisted transfer probability for the state 
\(\left|SS\right\rangle_1= \left|\psi^-\right\rangle_{AB} \otimes \left|\psi^-\right\rangle_{CD}\)
is unity. For the other options, viz. \(\left|SS\right\rangle_2= \left|\psi^-\right\rangle_{AC} \otimes \left|\psi^-\right\rangle_{BD}\) or 
\(\left|SS\right\rangle_3=\left|\psi^-\right\rangle_{AD} \otimes \left|\psi^-\right\rangle_{BC}\), it is zero. But the unassisted capacity will still be 
given by 
\({\cal C}_{ua}\left( \left|SS\right\rangle \right) = 1\). Using the method of entanglement swapping \cite{eita-swapping},
we have that the assisted capacity is unity for all the three options: 
\({\cal C}_{a}\left( \left|SS\right\rangle \right) = 1\). To see this, note that the assisted transfer probability is unity by construction
for the state \(\left|SS\right\rangle_1\), while entanglement swapping can to be employed to produce unit probabilities for the states
 \(\left|SS\right\rangle_2\) and \(\left|SS\right\rangle_3\).
By definition, the GGM of \(\left|SS\right\rangle\) is zero, while the GM of this state can be calculated to be \(3/4\).

\begin{figure}[h!]
\label{fig-chhobi-ek}
\begin{center}
\epsfig{figure= 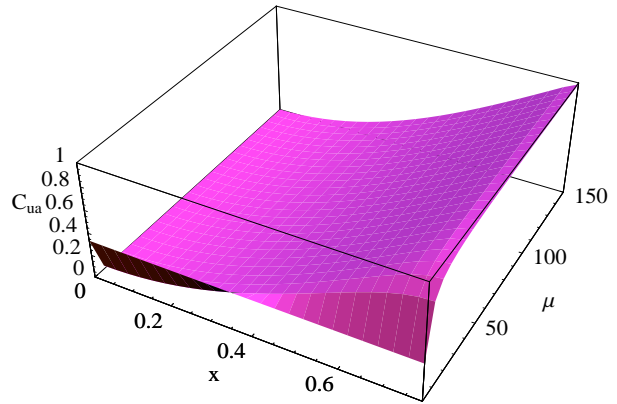, height=.27\textheight,width=0.45\textwidth}
\caption{
(Color online.) From resonating valence bond states to ferromagnets. The unassisted capacity (in ebits)
is plotted on the vertical axis against a base consisting of the measurement parameter \(x\) 
(for arbitrary projective measurements in the basis 
\(\{\cos x|0\rangle + \exp{(i\varphi)} \sin x|1\rangle, \exp{(-i\varphi)} \sin x|0\rangle - \cos x |1\rangle,\}\) at two parties), 
and the state parameter \(\mu\), for the family of four-qubit quantum states \(\left|\phi^{\mu}_{RF}\right\rangle\). (The base variables are dimensionless.)
The capacity for this strategy
can be readily read off from the figure for any \(\mu\). E.g. for the resonating valence bond state, situated at 
\(\mu=0\), the capacity is \(1/3\), while for 
the ferromagnetic state at \(\mu \rightarrow \infty\), the corresponding quantity is unity.
}
\end{center}
\end{figure}

\subsection{Case VI: RVB to Ferromagnets} The resonating-valence-bond state \cite{RVB}, apart from its significance 
in many-body physics, has 
potential applications in 
quantum information \cite{kitaev}. 
In our case, it can be expressed as 
 \(|\mbox{RVB}\rangle_{ABCD}=\big(\frac{1}{\sqrt{2}}(|01\rangle - |10\rangle)_{AB} \frac{1}{\sqrt{2}}(|01\rangle - |10\rangle)_{DC}
+ \frac{1}{\sqrt{2}}(|01\rangle - |10\rangle)_{AC} \frac{1}{\sqrt{2}}(|01\rangle - |10\rangle)_{DB}\big)/\sqrt{3} \), 
where  \(A\) and \(D\) are in sublattice 1, while \(B\) and \(C\) are 
in sublattice 2, of the bipartite lattice formed by \(A\), \(B\), \(C\), and \(D\), 
with the singlets being always directed from sublattice 1 to sublattice 2. The state can be rewritten as 
\(\left|\psi^{\mu}_{RF}\right\rangle_{ABCD}= (|0101\rangle + |1010\rangle + |0011\rangle + |1100\rangle - \mu |1001\rangle -\mu |0110\rangle)_{ABCD}/\sqrt{4+2\mu^2}\),
for \(\mu=2\). Interestingly, the state \(\left|\psi^{\mu}_{RF}\right\rangle\) is the ferromagnetic ground state (GHZ state of \texttt{Case I}) 
for \(\mu \rightarrow \infty\), and we will consider the state for the whole range \([2, \infty)\).
Certainly we have \({\cal C}_a \leq 1\), for all \(\mu\), but this bound can be attained by the following protocol. Suppose 
that \(A\) and \(B\) are allowed to share the additional 
singlet state resource. This implies that entangled measurements are allowed in the \(AB\) sector of the state \(\left|\psi^{\mu}_{RF}\right\rangle\). 
A measurement in the Bell basis \cite{eita-Bell-basis} at \(AB\) and subsequent local unitary transformations at \(C\) and \(D\), attains the bound 
for all \(\mu\). By explicit calculation, we find that \(p_{\mbox{\scriptsize{maxmin}}}^d \left( \left|\psi^{\mu}_{RF}\right\rangle \right)
= (\mu^2 - 2\mu +3)/(\mu^2 +2)\), so that by Proposition C2, \({\cal C}_{ua} \left(\left|\psi^{\mu}_{RF}\right\rangle\right) \leq (\mu^2 - 2\mu +3)/(\mu^2 +2)\). 
Measurements in the \(\{|+\rangle, |-\rangle\}\) basis by two parties, and local operations by the remaining parties, produces a lower bound:
\({\cal C}_{ua} \left(\left|\psi^{\mu}_{RF}\right\rangle\right) \geq (\mu^2 - 2\mu +2)/(\mu^2 +2)\).
For the RVB state, this reduces to \(1/3 \leq {\cal C}_{ua}(|\mbox{RVB}\rangle) \leq 1/2\). We have optimized over all projective 
measurements at two parties and all LOCC at the other two, for all \(\mu\), and the results are summarized in Fig. 1.
For the RVB state, the lower bound is the optimal one for the considered strategy. The generalized geometric measure of 
\(\left|\psi^{\mu}_{RF}\right\rangle\) is 
\((\mu+1)^2/(4+ 2\mu^2)\)
(by Proposition E1), so that for the RVB state, \({\cal E}= 1/4\). [\({\cal E}_G(|RVB\rangle) = 2/3\).]

\begin{figure}[h!]
\label{fig-chhobi-dui}
\begin{center}
\epsfig{figure= 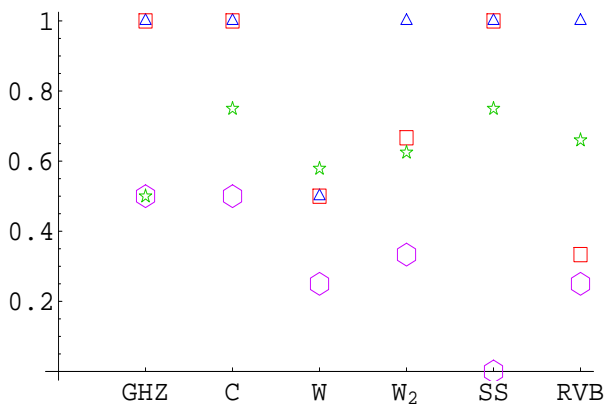, height=.23\textheight,width=0.45\textwidth}
\caption{
(Color online.) 
The capacities and the measures. The assisted capacities (blue triangles), unassisted capacities (red boxes), 
generalized geometric measures (pink hexagons), and geometric measures (green stars) for  a selection of 
four-party quantum states that are important from a quantum networks perspective. While the capacities are measured in ebits, the measures are 
dimensionless.
}
%
%
%
%
%
%
\end{center}
\end{figure}

\subsection{Bipartite versus multipartite revisited} 
Consider now the generalized GHZ (\texttt{Case I}) and RVB states (\(\left|\psi^{\mu}_{RF}\right\rangle\) for 
\(\mu=2\) in \texttt{Case VI}). Choosing \(\beta\) in the range \(1/4 < |\beta|^2 < 1/2\), we find that while the assisted capacity 
increases from generalized GHZ to RVB, the unassisted capacity actually decreases.  This therefore 
leaves us with no option to reconcile with the picture in the bipartite domain by using \emph{any} multipartite entanglement measure. 
 The multi-access channel capacities therefore presents a much richer picture than its 
bipartite variety. 
A similar situation arises if we compare the generalized GHZ states for \(\beta\) in the range \(1/3 < |\beta|^2 <1/2\) with the 
\(\left|\mbox{W}_2\right\rangle\)  state (\texttt{Case IV}).

It is plausible that the unassisted capacity for the RVB state is \(1/3\) (see \texttt{Case VI}). In that case, again such an irreconcilable situation
arises for the W (\texttt{Case III}) and RVB pair.  



The richness of the multiparty picture is further enforced by the other examples considered. 
In particular, the generalized GHZ and generalized W states reveal a situation where both the assisted and unassisted capacities are equal
(for certain choices of the parameters), while the GGM can still be different.
A synopsis of the whole picture is presented in Fig. 2.

\section{Conclusions}
Capacities of quantum channels corresponding to shared bipartite pure quantum states 
presents a relatively simple image, viz. the capacities are monotonically increasing 
with similar behavior for entanglement of the states.
Capacities of multi-access channels, however,  offers a much richer picture. 
Two such quantum capacities are defined and considered for paradigmatic multiparty quantum 
states, and compared against a measure of genuine multiparty entanglement. 
The quantum capacities are defined from the perspective of quantum computational networks. 

The measure of genuine multiparty entanglement, which 
we call the generalized geometric measure, is defined, its properties are explored. In particular, we find that 
it is possible to render it into a computable form for any multiparty quantum state of any dimension and of any number of parties. 

The investigation also points to the fact that at least for some multiparty situations, the additional singlet state does \emph{not} help to increase the capacity 
with respect to that in the unassisted case. This is in contrast to the bipartite case, where the singlet state is almost always the most important resource.

\acknowledgments
We acknowledge partial support from the Spanish MEC (TOQATA (FIS2008-00784)).

\end{document}